# Study of singular radius and surface boundary constraints in refractive beam shaper design

C. Liu[1,2], S. Zhang[1] [*]

[1]*Thomas Jefferson National Accelerator Facility, Newport News, Virginia 23606, USA*
[2] *MOE Key Laboratory of Heavy Ion Physics, Institute of Heavy Ion Physics, Peking University, 100871, China*

*[*]CorrespondingEmail:shukui@jlab.org*

**Abstract:** This paper presents analysis of important issues associated with the design of refractive laser beam shaping systems. The concept of "singular radius" is introduced along with solutions to minimize its adverse effect on shaper performance. In addition, the surface boundary constraint is discussed in detail. This study provides useful guidelines to circumvent possible design errors that would degrade the shaper quality or add undesired complication to the system.

©2008 Optical Society of America

## 1. Introduction

There is an increasing demand for laser beams with uniform intensity distribution for material processing, lithography, medical applications, laser printing, optical data storage, micromachining, isotope separation, optical processing, and other laboratory research [1]. The study of laser beam shaping has a long history and many papers on this subject can be found in the literature [1]. According to the chosen design method, beam shaping systems can be divided into three broad categories: aperture, integrator and field mapping. This paper presents an analysis of a refractive field mapping system. Many authors discuss how refractive shapers transform the laser beam intensity distribution [2-6]. A recent study shows that it is possible to shape an arbitrary beam profile into a desired one using only a single aspheric lens instead of the commonly used two-lens system [2-3].

An important issue associated with the popular refractive beam shaper has recently drawn considerable attention, that is, how to minimize the destructive effect of diffraction on the beam uniformity as it propagates through the distance needed for any practical application. Hoffnagle and Jefferson proposed redistributing the whole beam to a continuous roll-off beam profile [7~8]. Shealy and Hoffnagle considered output beam radiance distribution of four families of functions (super-Gaussian, flattened Gaussian, Fermi–Dirac, super-Lorentzian) and compared the propagation for each [9]. In this paper, we report a very important issue about the shaper design that, to the best of our knowledge, has never been discussed before. We refer to it as the "singularity" in the analytical solution of the shaper equations. In the following sections we discuss the meaning of the "singularity" and how to both avoid and benefit from this unexpected side effect. Finally, the "surface boundary constraint" will be discussed.

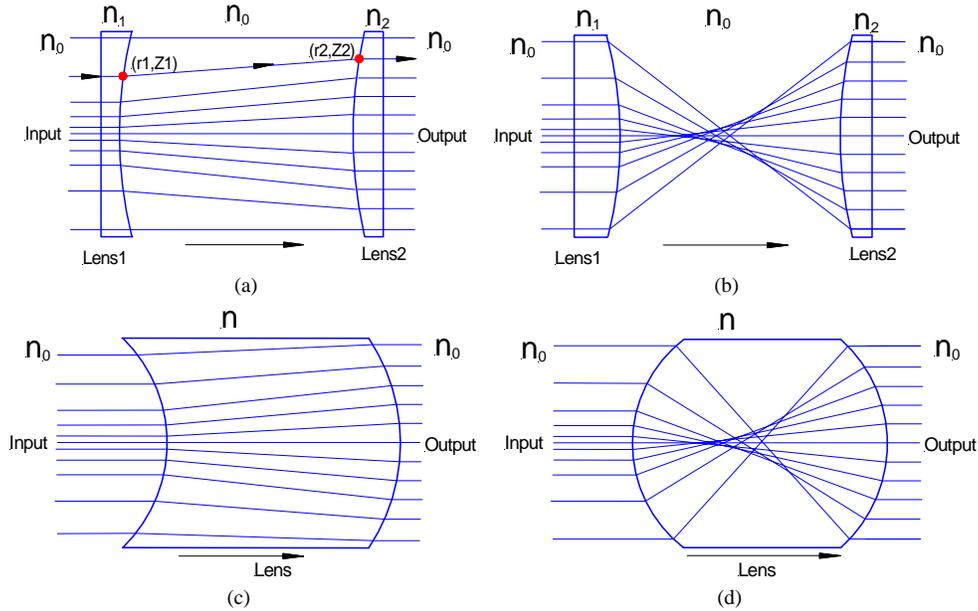

Fig. 1 Optical configurations of refractive laser beam shaper: (a) Galilean type,
(b) Keplerian shaper, (c) Single-lens Galilean shaper, (d) Single-lens Keplerian shaper

## 2. Shaper Design and "Singular Radius" Concept

The basic principle of laser beam shaping using a pair of aspheric lenses was proposed by Frieden [10] and Kreuzer [11], and later by Shealy and co-workers [12-13]. Fig. 1 shows an overview of the basic optical configurations of refractive beam shaper. For each case, a ray parallel to optical axis Z enters the first aspheric surface with arbitrary radius $r_1$, emerges from the second surface with radius $r_2$ and is parallel to Z axis again. The terms $Z_1(r)$ and $Z_2(r)$ represent the first (front) and second (rear) surface profiles, respectively, $n$ is the index of refraction, and $s$ is the distance between the vertex of the two surfaces. These beam shapers must all meet two requirements: (1) energy conservation, the total beam energy remains constant from the input to the output, and (2) equal optical path, all rays passing through the shaper from input pupil to output pupil have zero optical path difference (OPD). These conditions guarantee the shaped beam is perfectly collimated at the output. Applying Snell's Law at each lens surface, surface profiles can be defined by the following relationship between $r_1$, $Z_1(r)$, $r_2$ and $Z_2(r)$:

Type-1: Conventional two-lens Galilean shaper:

$$Z_1(r) = \int_0^r [(n^2-1) + (\frac{(n-1)s}{r_2-r_1})^2]^{-1/2} dr_1 \tag{1a}$$

$$Z_2(r) = \int_0^r [(n^2-1) + (\frac{(n-1)s}{r_2-r_1})^2]^{-1/2} dr_2 \tag{1b}$$

Type-2: Two-lens Keplerian shaper:

$$Z_1(r) = \int_0^r [(n^2-1) + (\frac{(n-1)s}{r_2+r_1})^2]^{-1/2} dr_1 \tag{2a}$$

$$Z_2(r) = \int_0^r [(n^2-1) + (\frac{(n-1)s}{r_2+r_1})^2]^{-1/2} dr_2 \tag{2b}$$

Type-3: Single-lens Galilean shaper:

$$Z_1(r) = \int_0^r n[-(n^2-1) + (\frac{(n-1)s}{r_2-r_1})^2]^{-1/2} dr_1 \tag{3a}$$

$$Z_2(r) = \int_0^r n[-(n^2-1) + (\frac{(n-1)s}{r_2-r_1})^2]^{-1/2} dr_2 \tag{3b}$$

Type-4: Single-lens Keplerian shaper:

$$Z_1(r) = \int_0^r n[-(n^2-1) + (\frac{(n-1)s}{r_2+r_1})^2]^{-1/2} dr_1 \tag{4a}$$

$$Z_2(r) = \int_0^r n[-(n^2-1) + (\frac{(n-1)s}{r_2+r_1})^2]^{-1/2} dr_2 \tag{4b}$$

It is assumed the input beam radiance is a normalized Gaussian profile $f(r)$ and output beam is a super-Gaussian $g(r)$. The relationship between $r_1$ and $r_2$ is determined by the energy conservation condition. To avoid diffraction effects, the shaper radius was designed to be 5.7mm so that the radiance of the input Gaussian beam at the edge of the shaper is $e^{-7}$ of the peak value. In our design, input beam size was set with w=3mm. The parameter R, a length scale approximately the radius of a flattop for P>2, is 4mm, together with the dimensionless order parameter P=8, define the output super-Gaussian radiance profile:

$$g(r) = g_0 \exp(-2(r/R)^P) \tag{5}$$

$$g_0 = \frac{2^{2/P} P}{2\pi R^2 \Gamma(2/P)} \tag{6}$$

After straight-forward analytical derivation using the energy conservation requirement a second critical design equation is obtained,

$$r_1 = h(r_2) = \sqrt{-\frac{9}{2}\ln(1 - 2\pi \int_0^{r_2} g(r)r dr)} \tag{7}$$

The above equations define the specific aspheric surface profiles for the corresponding shaping system. It needs to be pointed out that Eqs. 1a, 1b are from Ref. 11, Eqs. 5, 6 are from Ref. 9 and we further derived Eq. 2a through Eq. 4b and Eq. 7 for different shaper configurations. Compared with the differential equations used in traditional shaper design these equations are quite lucid and could be solved analytically for special cases, for example, the Gaussian to super-Gaussian transformation. The use of these equations in a design will be presented later. It is worthwhile to mention that Eq. 7 applies to all four design types listed above.

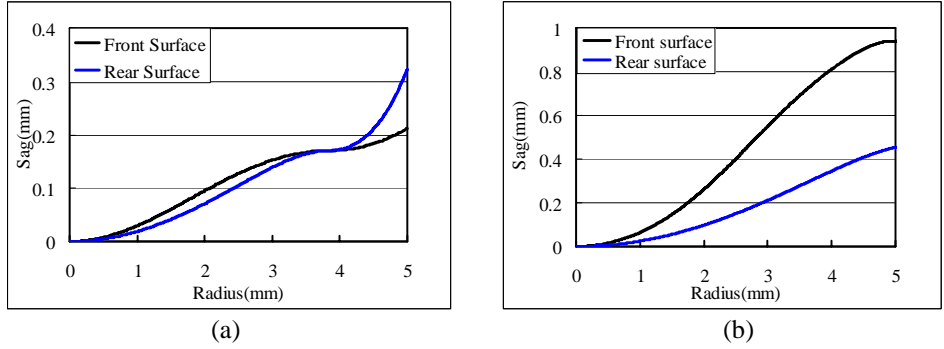

Fig. 2 Sag curves for the two single-lens beam shapers, (a) R=4mm, and (b) R=5mm.

Consider, as an example, a Type-3 shaping system with 30mm lens thickness of BK7 glass and design wavelength of 532nm. The surface profiles for two different design output beam sizes characterized by R=4mm and R=5mm are shown in Fig. 2(a) and 2(b); alongside simulation results for R=4mm using a ray tracing program (ZEMAX). Both designs produce high quality flat-top patterns. Fig. 3 shows the radiance distributions for R=4mm. Note in Fig. 2a there is a turning point near $r$ =3.8mm where, for the R=4mm curve, the first derivative $dZ/dr = 0$. In this case, both the front and rear surfaces are no longer monotonic in terms of optical element smoothness. However, the curve for R=5mm is monotonic and quite normal. The occurrence of a tuning point is highly undesirable because it incurs unnecessary complication, adding particular difficulty in data processing and optical fabrication. Although the turning point is near the edge of the aspherical surface, it is not a location at which either the input or the output radiance may be ignored. For the case R=4mm, the intensity at the turning point ($r$ =3.8mm) of the input beam is about 4% of the peak value, yet over 26.5% for the output beam! For convenience, we define the radius where the turning point appears as the "singular radius" and next discuss its interpretation.

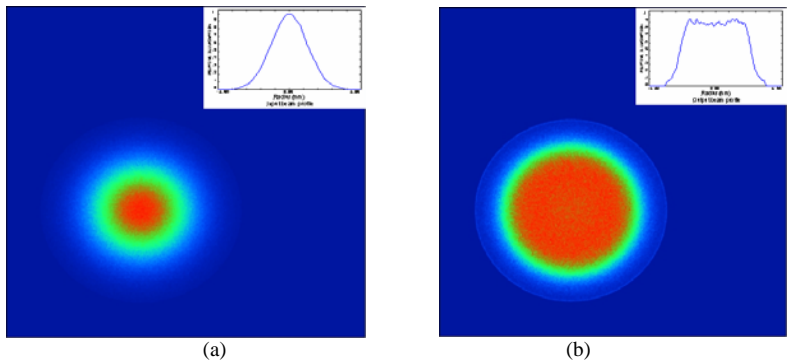

Fig. 3 Input and output beam radiance distributions for R=4mm in Fig. 2(a),
(a) Input profile, (b) Output profile

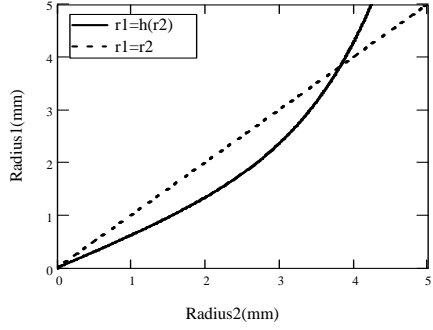 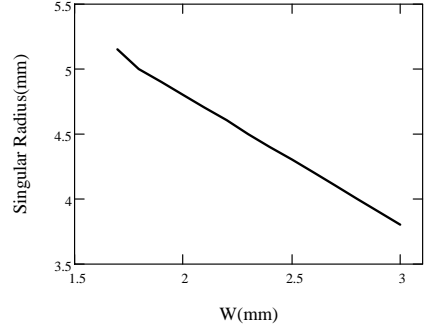

Fig. 4 Curve showing the relation between $r_1$ and $r_2$.

Fig. 5 Singular radius changes with initial beam size

It is not difficult to understand the nature of the singular radius. For example, consider the Type-1 and Type-3 shapers. The function of a beam shaper is to uniformly distribute the input beam energy over the designated output surface area. For rays near the optical axis, their radiance is higher than those near the surface edge. Consequently, rays at smaller radius $r_1$ on the front surface tend to defocus more than those at larger $r_1$, ending at with larger value of $r_2$ on the rear surface. This situation is reversed for those rays near the surface edge since the shaped flat-top radiance should be between the center peak and the lower wing of the original Gaussian input radiance. It is easy to imagine there must be a point where the input ray will pass straight through the optical material and reach the rear surface without any focusing or defocusing. In other words, the incident ray from $r_1$ will travel parallel to the optical axis, indicating $r_1 = r_2$. In comparison, rays towards the edge are bent at a larger angle for Type-2 and Type-4 shapers, so there is no other straight ray except the center ray on optical axis where $r_1 = r_2 = 0$. A straight-forward way to predict the presence of the singular radius is to make a plot like that shown in Fig. 4, where the singular radius is obtained at the intersection point of a line ($r_1 = r_2$) and the curve defined by Eq. 7. Analysis of the relationship between singular radius and other parameters of the shaping system are presented below. We show that it is possible to eliminate the singular radius from Type-1 and Type-3 designs.

## 3. Analysis

For the sake of simplicity and consistency, we change only one parameter in the following discussion regarding a Type-3 shaper. In addition, we focus on the singular radius at the rear surface because the singular radii are always equal.

3.1 Initial beam size (w)

The dependence of the singular radius with initial beam size in Fig. 5 demonstrates that the smaller input beam size results in a larger singular radius. When w is decreased to about 1.4mm, singular radius moves far from our desired output beam size range. As explained above, the smaller input beam tends to defocus more in order to redistribute the total energy onto a given surface area. When the area ratio between the input and output beams reaches a threshold, even rays near the edge of the input beam will not focus. The consequence is that the singular radius cannot occur. This indicates an important criterion for designing shaper systems: the use of a smaller input beam helps to eliminate the singular radius for a given output aperture.

3.2 Lens thickness ($s$)

The singular radius hardly changes as the lens thickness varies, however, the surface profiles show significant differences, as demonstrated in Fig. 6. The curve becomes more monotonic as the lens thickness increases. The reason the singular radius does not change with $s$ is clear: singularity occurs when $r_1$ equals $r_2$, and is solely determined by Eq. 7 which has no dependency on lens thickness.

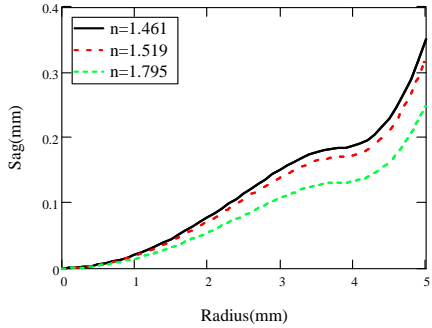

Fig. 7 Sag curves of rear surface for different index of refractive. $S$ =30mm.

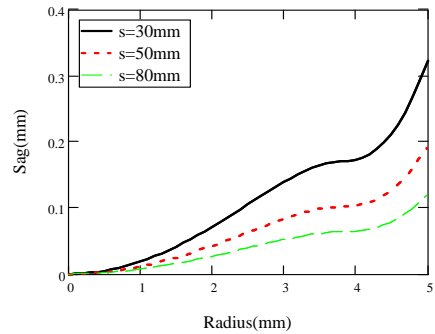

Fig. 6 Sag curves of rear surface for different lens thickness

3.3 Index of refraction ($n$)

The dependency of the singular radius on material refraction index presents a very similar picture as that for lens thickness. The reason is the same as explained in 2). From Fig. 7 we can see in this case the change of $n$ won't make the surface profile get closer to monotonic. With larger $n$, the sag value does decease, though. This helps to facilitate the optical fabrication.

3.4 Output beam parameters P and R

Fortuitously, the super-Gaussian order parameter (P) does not influence the singular radius. In this specific design, the singular radius ($r$ =3.8mm) is somewhat smaller than the length scale parameter (R=4mm) which means the singular radius is along the ramp to the flattop profile. This is true for all design parameters. It is true also that the ramp of the output beam does not move much with order parameter however the slope does change. Consequently, the singular radius does not change significantly with P over a range of interest (P>8 for good uniformity and P<20 for diffraction suppression). However, as shown in Fig. 8, it would be extremely difficult to design a shaper which preserves a small beam while flattening the radiance profile, and difficult to constrict the beam while shaping.

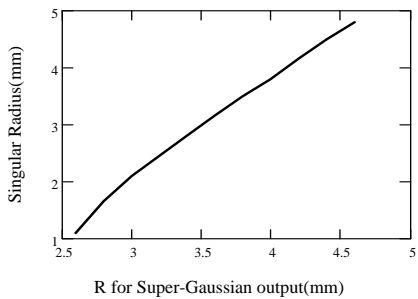

Fig. 8 Singular radius changes with output beam parameter R

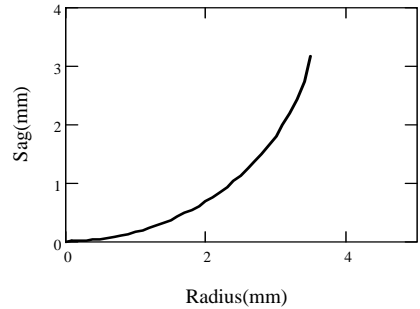

Fig. 9 Surface profile showing the boundary constraint for type-4 shaper

In a similar manner, the singular radii for Type-1 and Type-3 shaping system used to create other output beam profiles—for example, Fermi-Dirac, Flattened-Gaussian and Super-Lorentzian, even from an arbitrary input profile to an arbitrary output profile, can be treated.

## 4. Surface boundary

Although neither Type-2 nor Type-4 shapers have singular radii, there is another concern associated with Type-3 and Type-4 configurations. The slope of the lens surface (dr/dZ) may become very small or even reach zero at a radius at which the integrand in (3) and (4) goes to infinity. At the same location, the surface becomes parallel with the optical axis. This is equivalent to the incident angle reaching the critical angle for total reflection. The sag for a Type-4 shaper that converts the input beam (W=3mm) into a super-Gaussian output beam (P=8, R=6mm) is shown in Fig. 9, the lens separation is now 15mm. As we can see, the surface ends at radius less than 4mm. Therefore, there is no room for extending the shaper output aperture in order to minimize diffraction effects which may severely affect the beam pattern within a certain distance. This leads to a very important boundary issue for Type-3 and Type-4 shapers. We discuss the two types separately:

1) Type-3

This type of shaper can be divided into two configurations: one a diverging shaper with concave front surface and convex rear surface, the other a converging shaper with convex front and concave rear. Differences between corresponding $r_1$ and $r_2$ of the converging shaper could be big enough to meet the following condition; however, it is not possible for the diverging case. The surface ends as the integrand in Eq. 3 goes to infinity,

$$r_1 - r_2 = \sqrt{\frac{n-1}{n+1}}s \qquad (8)$$

Clearly, with short lens thickness $s$ the surface terminates at a radius far enough from the optical axis, and it will go further when s increases, which is revealed by Fig. 10.

2) Type-4

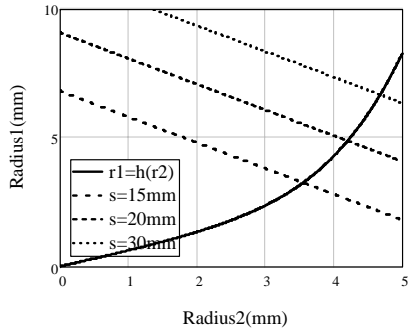

Fig. 11 Shaper surface boundary changes with lens thickness for type-4 shaper

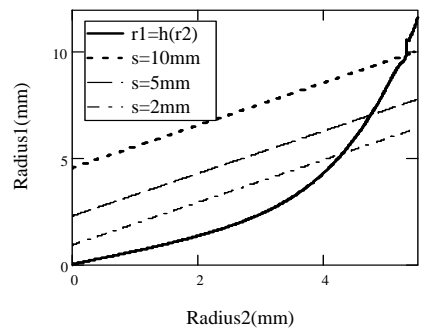

Fig. 10 Dependency of type-3 beam shaper surface boundary on different lens thickness

In this case the boundary equation will be

$$r_1 + r_2 = \sqrt{\frac{n-1}{n+1}} s \tag{9}$$

The calculation is shown in Fig. 11. As lens thickness $s$ is increased the surface extends further outward. For $s$ <15mm, the constraint is fairly rigorous for the design. It is easy to see that more attention needs to be taken when designing Type-4 than Type-3, because a Type-4 shaper inherently has faster bending surface profile. As we know, short spacing $s$ is desirable for ultra short pulse laser beam shaping and a larger aperture is much easier to fabricate and desirable for diffraction suppression. Due to the boundary constraint it is necessary to analyze the compromise between spacing and aperture in order to optimize a specific design.

### 5. Summary

We have introduced the concept of "singular radius" in this paper along with detailed discussions about the surface boundary constraint on the refractive shaper design. By analyzing the issues and influence by other parameters we put forward practical measures to eliminate such problems from the design process. To the best of our knowledge these issues and the study presented in this paper have not been reported elsewhere in the literatures. We believe this paper provides very useful guidelines against possible design errors that could otherwise degrade shaper performance or add undesired complication into the system.

### Acknowledgement

This work is supported by the Office of Naval Research, the Joint Technology Office, the Commonwealth of Virginia, the Air Force Research Laboratory, and by DOE Contract DE-AC05-060R23177. C. Liu also acknowledges the support from China Scholarship Council.